\documentstyle[epsf,12pt]{article}
\begin{document}
\title{
 A self--similar dynamics in viscous spheres}
\author{}
\maketitle
\begin{center}
         {\large W. Barreto}%
         \footnote[1]{Laboratorio de F\'{\i}sica Te\'orica,
         Departamento de F\'{\i}sica,
         Escuela de Ciencias, N\'ucleo de Sucre,
         Universidad de Oriente,
         Cuman\'a, Venezuela}$^{,}$%
         \footnote[2]{Centro de Astrof\'{\i}sica
         Te\'orica,
         Universidad de los Andes,
         M\'erida, Venezuela.}
         {\large, J. Ovalle}%
         \footnote[3]{Departamento de F\'{\i}sica,
         Escuela de F\'{\i}sica y Matematicas,
         Facultad de Ciencias,Universidad Central de Venezuela,
         Caracas, Venezuela.}$^{,}$%
         \footnote[4]{Departamento de F\'{\i}sica,
         Universidad Sim\'on Bol\'\i var,
         Caracas, Venezuela.}
         {\large and B. Rodr\'\i guez}$^{1}$
\end{center}
\begin{abstract}
We study the evolution of radiating and viscous fluid spheres 
assuming an additional homothetic symmetry on the spherically simmetric
 space--time. 
 We match a very simple solution to the symmetry
equations with
 the exterior one (Vaidya).
 We then obtain 
 a system of two ordinary differential equations which
 rule the dynamics, and find a self--similar collapse which is shear--free
 and with a barotropic equation of state. Considering a huge set of
 initial self--similar dynamics states, we work out a model with an acceptable
 physical behavior.
\end{abstract}
\section{Introduction}
 Often many authors assume spherical symmetry and perfect fluid approx\-i\-ma\-tion
 to face the problem of self--gravitating and collapsing distributions
 of matter. Also, they use extensively 
 progressive waves or similarity solutions 
 (see \cite{ct71,op90} and references therein). If the fluid is perfect
 the only equation of state compatible with self--similar fluids is the 
 barotropic one \cite{op90}. The present paper concerns in part with the
 validness of the barotropic equation of state for a viscous and radiating
 fluid sphere.

In general, there are two types of self--similar space--times depending
 on whether they are invariant or not under scale transformations. 
Scale--free self--similar solutions are the similarity solutions of type
 one and the resulting space--time admits homothetic Killing vectors.
 Type two similarity solutions are not invariant under the simple scaling
 group \cite{hew83}--\cite{c97}. The self--similar symmetry has been 
 reported to characterize
 these two types of self--similar space--times \cite{p93}.

 Spherically symmetry and homothetic space--times show naked
 singularities. Assumption of similarity rather than spherical
 symmetry is crucial in determining the nature of the singularity
 in any gravitationally collapsing configuration \cite{hp91,sc96}. 
 So far, self--similar space--times have been studied mainly in 
 cosmological contexts
 \cite{bh78}--\cite{cv94}.

Considering that the perfect fluid approximation is
 likely to fail, at least in some stages of stellar collapse, 
 in this paper we study radiating and viscous fluid spheres.
 Specifically,
 we have been concerned
 with the radiative shear viscosity and its effect on the gravitational
 collapse \cite{hjb89}--\cite{b93}.  We do not consider here the temperature
 profiles to determine which processes can take place during the collapse.
 For this purpose, transport equations have been proposed to avoid
 pathological
 behaviors (see for instance \cite{m96} and references therein).
 The motivation of this work was a recent study of radiating and
 dissi\-pative spheres \cite{bc95}.
 We assume an additional symmetry (homothetic motion)
 within the viscous fluid sphere without heat flow in the streaming out
 limit.
 
 The organization of this paper is the following. Section 2 shows the field
 equations,
 the junction conditions and the surface equations.
 In section 3 we write the homothetic
 motion equations in a convenient form.  We propose a very simple solution 
 in section 4 to work out some models. Finally, in section 5, we draw
 conclusions. 

\section{Dynamics and matching}
\subsection{Field equations}
~~~~To write the Einstein field equations we use the line element in Schwarzs\-
child--like coordinates
\begin{equation}
ds^2=e^\nu dt^2-e^\lambda dr^2-r^2\left( d\theta ^2+\sin
{}^2\theta d\phi ^2\right). \label{eq:metric} 
\end{equation}
where $\nu = \nu(t,r)$ and $\lambda = \lambda(t,r)$, with
 $(t,r,\theta,\phi)\equiv(0,1,2,3)$.

In order to get physical input we introduce the
 Minkowski coordinates $(\tau,x,y,z)$ by \cite{b64}
\begin{equation}
d\tau=e^{\nu /2}dt,\,  
dx=e^{\lambda /2}dr,\,  
dy=rd\theta,\, 
dz=r \sin \theta d\phi,\label{eq:local}
\end{equation}
In these expressions $\nu$ and $\lambda$ are constants, because they
 have only local values. 

Next we assume that, for an observer moving relative to these coordinates
 with velocity $\omega$ in the radial ($x$) direction, the space contains
\begin{itemize}
\item a viscous fluid of density $\rho$, pressure $\hat p$, effective
 bulk pressure $p_\zeta$ and effective shear pressure $p_\eta$, and
\item unpolarized radiation of energy density $\hat \epsilon$.
\end{itemize}

For this moving observer, the covariant energy tensor in Minkowski
 coordinates is thus
 
\begin{equation}
\left(
\begin{array}{cccc}
\rho+\hat \epsilon & -\hat \epsilon & 0 & 0 \\
-\hat \epsilon & \hat p +\hat \epsilon - p_\zeta - 2 p_\eta& 0 & 0 \\
0 & 0 & \hat p - p_\zeta + p_\eta & 0 \\
0 & 0 & 0 & \hat p - p_\zeta + p_\eta
\end{array}
\right)
\end{equation}

Note that from (\ref{eq:local}) the velocity of matter in the Schwarzschild
 coordinates is
\begin{equation}
\frac{dr}{dt} = \omega e^{(\nu-\lambda)/2} \label{eq:velocity}
\end{equation}

Now, by means of a Lorentz boost and defining  
 $\tilde p \equiv \hat p - p_\zeta$,
 $p_r \equiv \tilde p - 2 p_\eta$,
 $p_t \equiv \tilde p + p_\eta$ and
 $\epsilon \equiv \hat \epsilon(1+\omega)/(1-\omega)$ we write 
 the field equations in relativistic units ($G=c=1$) as follows: 
\begin{equation}
\frac{\rho + p_r \omega^2}{1-\omega ^2} + \epsilon =
\frac{1}{8\pi r}\left[\frac{1}{r} - 
e^{-\lambda}\left(\frac 1{r}-\lambda_{,r}\right)\right] \label{eq:ee1}
\end{equation}

\begin{equation}
\frac{p_r + \rho \omega^2}{1-\omega ^2} + \epsilon =
\frac{1}{8\pi r}\left[
e^{-\lambda}\left(\frac 1{r}+\nu_{,r}\right) - \frac{1}{r}\right] \label{eq:ee2}
\end{equation}

\begin{eqnarray}
p_t = \frac{1}{32\pi} \{ e^{-\lambda}[ 2\nu_{,rr}+\nu_{,r}^2
-\lambda_{,r}\nu_{,r} + \frac{2}{r}
(\nu_{,r}-\lambda_{,r}) ] - \nonumber \\ \nonumber \\
e^{-\nu}[ 2\lambda _{,tt}+\lambda_{,t}(\lambda_{,t}-\nu_{,t}) ] \} \label{eq:ee3}
\end{eqnarray}

\begin{equation}
(\rho + p_r)\frac{\omega}{1-\omega^2} + \epsilon = 
-\frac{\lambda_{,t}}{8\pi r}e^{-\frac 12(\nu+\lambda)} \label{eq:ee4}
\end{equation}
where the comma (,) represents partial differentiation with 
 respect to the indicated
 coordinate.
Equations (\ref{eq:ee1})--(\ref{eq:ee4}) are formally the same as for
 an anisotropic fluid
 in the streaming out approximation.

At this point, for the sake of completeness, we write the effective
 viscous pressures in terms of
 the bulk viscosity $\zeta$, the volume expansion $\Theta$, the shear
 viscosity $\eta$ and the scalar shear $\sigma$ \cite{b93} 
\begin{equation}
p_\zeta = \zeta \Theta
\end{equation}
\begin{equation}
p_\eta = \frac{2}{\sqrt{3}} \eta \sigma
\end{equation}
where
\begin{equation}
\Theta = \frac{1}{(1-\omega^2)^{1/2}}
\left[e^{-\nu/2}\left(
\frac{\lambda_{,t}}{2}+
\frac{\omega\omega_{,t}}{1-\omega^2}
\right)+e^{-\lambda /2}
\left(\frac{\nu_{,r}}{2}\omega+\frac{\omega_{,r}}{1-\omega^2}+\frac{2\omega}
{r} \right) \right]
\end{equation}
and
\begin{equation}
\sigma =\sqrt{3}\left( \frac \Theta 3-\frac{e^{-\lambda /2}}r\frac
\omega {\sqrt{1-\omega ^2}}\right)
\end{equation}

We have four field equations for six physical variables ($\rho$, $p$,
 $\epsilon$, $\omega$, $\zeta$ and $\eta$) and two geometrical
 variables ($\nu$ and $\lambda$). Obviously, we require additional
 assumptions to handle the problem consistently. First, however, we 
 discuss the matching with the exterior solution and the surface
 equations that govern the dynamics.

\subsection{Junction conditions}
~~~~We describe the exterior space--time by the Vaidya metric 
\begin{equation}
ds^2=\left( 1-\frac{2{\cal M}(u)}R\right) du^2+2dudR-R^2\left( d\theta
^2+\sin^2\theta d\phi^2 \right)
\end{equation}
where $u$ is a time--like coordinate so that $u=$ constant represents,
 asymptotically,  null
 cones open to the future and R is a null coordinate ($g_{RR}=0$). 
 The relationship between the coordinates ($t$,$r$,$\theta$,$\phi$)
 and ($u$,$R$,$\theta$,$\phi$) is
\begin{equation}
u=t-r-2{\cal M}\ln \left( \frac r{2{\cal M}}-1\right), R=r
\end{equation}

The exterior and interior solutions are separated by the surface $r=a(t)$.
 To match both regions on this surface we require the Darmois junction
 conditions. 
 Thus, demanding the continuity of the first fundamental form, we obtain
\begin{equation}
e^{-\lambda_a}=1-\frac{2{\cal M}}{R_a} \label{eq:ffa}
\end{equation}
and 
\begin{equation}
\nu_a = -\lambda_a \label{eq:ffb}
\end{equation}
From now on the subscript $a$ indicates that the quantity is evaluated 
 at the surface.
 Now,
 instead of writing the junction conditions as usual, we demand the continuity
 of the first fundamental form and the continuity of the 
 independent components of the energy--momentum flow. This last condition
 guarantees absence of singular behaviors on the
 surface. It is easy to check that \cite{b93,h96} 
\begin{equation}
\hat p_a = p_{\zeta _a} + 2 p_{\eta _a} \label{eq:boundary}
\end{equation}
which expresses the discontinuity of the radial pressure in presence of
 viscous processes.

\subsection{Surface equations}
~~~~To write the surface equations we introduce the mass function $m$
 by means of
\begin{equation}
e^{-\lambda (r,t)}=1-2 m(r,t)/r \label{eq:masa}
\end{equation}
Substituting (\ref{eq:masa}) into (\ref{eq:ee1}) and (\ref{eq:ee4})
 we obtain, after some arrangements, 
\begin{equation}
\frac{dm}{dt}=-4\pi r^2\left[\frac{dr}{dt}p_r
+\epsilon (1-\omega )(1-2m/r)^{1/2}e^{\nu /2} \right] \label{eq:energy}
\end{equation}
This equation shows the energetics across the moving boundary of the 
fluid sphere. Evaluating (\ref{eq:energy}) at the surface and using the boundary
 condition (\ref{eq:boundary}) (which is equivalent to $p_{r_a}=0$), the
 energy loss is given by   
\begin{equation}
\dot m_a =-4\pi a^2 \epsilon _a (1 - 2m_a/a) (1-\omega_a)
\end{equation}
Hereafter overdot indicates $d/dt$.
 The evolution of the boundary is governed by equation (\ref{eq:velocity})
 evaluated at the surface
\begin{equation}
\dot a=(1-2m_a/a)\omega _a
\end{equation}
Scaling the total mass $m_a$, the radius $a$ and
 the time--like coordinate by the initial mass $m_a(t=0)\equiv m_a(0)$,
$$ A\equiv a/m_a(0), \, M\equiv m_a/m_a(0), \, t/m_a(0) \rightarrow t$$
and defining
\begin{equation}
F\equiv 1-\frac{2M}A
\end{equation}
\begin{equation}
\Omega \equiv \omega _a
\end{equation}
\begin{equation}
E\equiv 4\pi a^2\epsilon _a(1-\Omega)
\end{equation}
the surface equations can be written as
\begin{equation}
\dot A=F\Omega \label{eq:first}
\end{equation}
\begin{equation}
\dot F=\frac FA\left[(1-F)\Omega +2E\right] \label{eq:second}
\end{equation}
Equations (\ref{eq:first}) and (\ref{eq:second}) are general
 within spherical symmetry. We need a third surface equation to specify
 the dynamics completely for any set of initial conditions and a given
 luminosity profile $E(t)$. For this purpose we can use equation (\ref{eq:ee3})
 or appeal to the conservation equation $T_{1;\mu }^\mu=0$ evaluated at the
 surface. But we follow here another route, that is, we assume that the 
 space--time admits a one--parameter group of homothetic motion generated
 by a homothetic Killing vector orthogonal to the four--velocity.
 These assumptions introduce some restrictions on the surface equations
 as is shown in the next section.
 
\section{Homothetic motion}
~~~~We assume that the spherically symmetric space--time within the fluid admits a one--parameter
 group of homothetic motions.  In general, a global vector field
 $\xi$ on the manifold is called homothetic if $\pounds _{\bf \xi}{\bf g}
 = 2 n {\bf g}$ holds on a local chart, where $n$ is a constant on the
 manifold, and $\pounds$ denotes the Lie derivative operator. If $n 
 \neq 0$, $\xi$ is called proper homothetic and it can always
 be scaled so to have $n = 1$; if $n = 0$ the $\xi$ is a Killing
 vector on the manifold \cite{h88}--\cite{cms94}. So, after a constant
 rescaling we write
 
\begin{equation}
\pounds _{\bf \xi}{\bf g}=2{\bf g} \label{eq:homothetic}
\end{equation}
where the vector field {\bf $\xi$} has the general form
\begin{equation}
{\bf \xi} =\Lambda (r,t)\partial _t +\Gamma (r,t)\partial _r
\end{equation}
After simple manipulations we obtain from (\ref{eq:homothetic})
\begin{equation}
\Gamma = r
\end{equation}
\begin{equation}
\Lambda_{,r} = 0
\end{equation}
\begin{equation}
\Lambda m_{,t} +\Gamma m_{,r} = m \label{eq:se3}
\end{equation}
\begin{equation}
\Lambda \nu_{,t} + \Gamma \nu_{,r} + 2 \dot \Lambda = 2 \label{eq:se4}
\end{equation}
We further assume that the four--velocity is orthogonal to the orbit of the 
 group
\begin{equation}
\omega=\frac{\Lambda}{r}e^{(\nu-\lambda)/2}
\end{equation}
Thus we obtain a connection between the time--like component
 of the homothetic Killing vector and the surface variables,
\begin{equation}
\Lambda (t)=\frac{a\Omega }F \label{eq:temporal}
\end{equation}
Now, expanding $\nu$ near the surface, using (\ref{eq:ffa}), (\ref{eq:ffb}),
 (\ref{eq:temporal}), and evaluating at $r=a$ the equations (\ref{eq:ee1}), 
 (\ref{eq:ee4}), (\ref{eq:se3}) and (\ref{eq:se4}), after straightforward
 manipulations we find the surface equation
\begin{equation}
\dot \Omega =\frac{(1-\Omega ^2)}{2A}\left( 3F-1-2E\right) \label{eq:omegadot}
\end{equation}
 From now on we disregard the bulk effective pressure to promote 
 algebraic consistence.
\section{Modeling}
~~~~In order to work out models we define the self--similar variables
\begin{equation}
X=\frac{m}r
\end{equation}
and
\begin{equation}
Y=\frac{\Lambda}r{e^{\nu /2}}
\end{equation}
Thus, equations (\ref{eq:se3}) and (\ref{eq:se4}) read
\begin{equation}
\Lambda X_{,t}+ rX_{,r}=0
\end{equation}
and
\begin{equation}
\Lambda Y_{,t}+ r Y_{,r} = 0
\end{equation}
In general these equations have solutions $X=X(\varsigma)$ and $Y=Y(\varsigma)$
, where $\varsigma$ is
\begin{equation}
\varsigma =re^{-\int dt/\Lambda }
\end{equation}
We propose the specific solutions
\begin{equation}
X=C_1\varsigma ^k \label{eq:X}
\end{equation}
and
\begin{equation}
Y=C_2\varsigma ^l \label{eq:Y}
\end{equation}
where $C_1$, $C_2$, $k$  and $l$ are constants.

Solutions (\ref{eq:X}) and (\ref{eq:Y}) are restricted by (\ref{eq:ffa})
 and (\ref{eq:ffb}). Therefore  the
 geometrical variables are  
\begin{equation}
m=m_a\left( \frac r{a}\right) ^{k+1} \label{eq:mass}
\end{equation}
\begin{equation}
e^\nu =F\left( \frac ra\right) ^{2(l+1)} \label{eq:nu}
\end{equation}
In order to get the unique luminosity
\begin{equation}
E=\frac{1}{2}[F(k+2l+3)-(k+1)] \label{eq:luminosity}
\end{equation}
we use equations (\ref{eq:ee1}), (\ref{eq:ee2}), (\ref{eq:mass}) and
 (\ref{eq:nu}) together with the boundary conditions 
 (\ref{eq:ffa}), (\ref{eq:ffb}) and (\ref{eq:boundary}) to find  
\begin{equation}
\Omega =\frac{2E \pm Z}{2(k+1)(F-1)} \label{eq:omega}
\end{equation}
where
\begin{eqnarray}
Z=[F^2(5k^2+4kl+10k+4l^2+12l+9) \nonumber \\ \nonumber \\
 -2F(k+1)(5k+2l+3)+5k^2+6k+1]^{\frac 12}
\end{eqnarray}
Note that ``$+$'' in the numerator of (\ref{eq:omega}) represents the
 collapsing solution and
 ``$-$'' an expanding one. We
 consider here only $\Omega^+$ situations.

Now, combining equations (\ref{eq:omegadot}) and (\ref{eq:omega})
 we obtain an equation $f(F,k,l)=0$, which is too lengthy to present
 here, but which permits us to model different situations. The first one is 
 the shear--free and self--similar collapse for which $k=l=0$, 
 $m/a \approx 0.3096$ ($m$ and $a$ are linear with time)
 and $\tilde p = 0$ at any space--time point. 
 The second possibility appears upon solving for $l=l(F(t=0),k)$ and includes
 the previous case. For $k \neq 0$ we obtain shearing models but 
 the homothetic symmetry is broken for $t > 0$. 

\begin{figure}
\centerline{\epsfxsize=6in\epsfbox{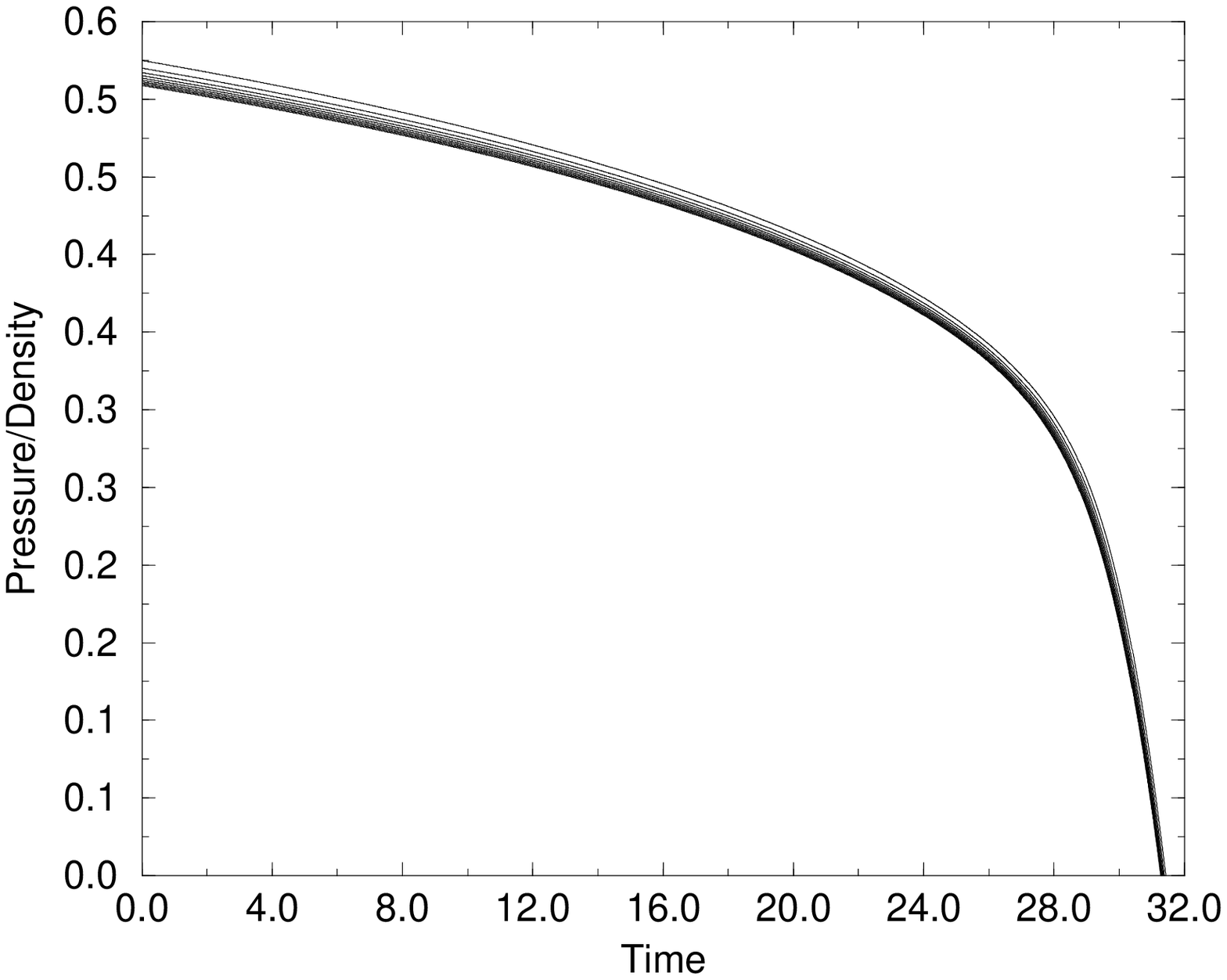}}
\caption{$\tilde p/ \rho$ as a function of time, for different values of 
 $r/a$: 0.1 (uppermost curve), 0.2, 0.3, 0.4, 0.5, 0.6, 0.7, 0.8, 0.9 and 1.0
 (lowermost curve).}
\label{fig:poverho}
\end{figure}

\begin{figure}
\centerline{\epsfxsize=6in\epsfbox{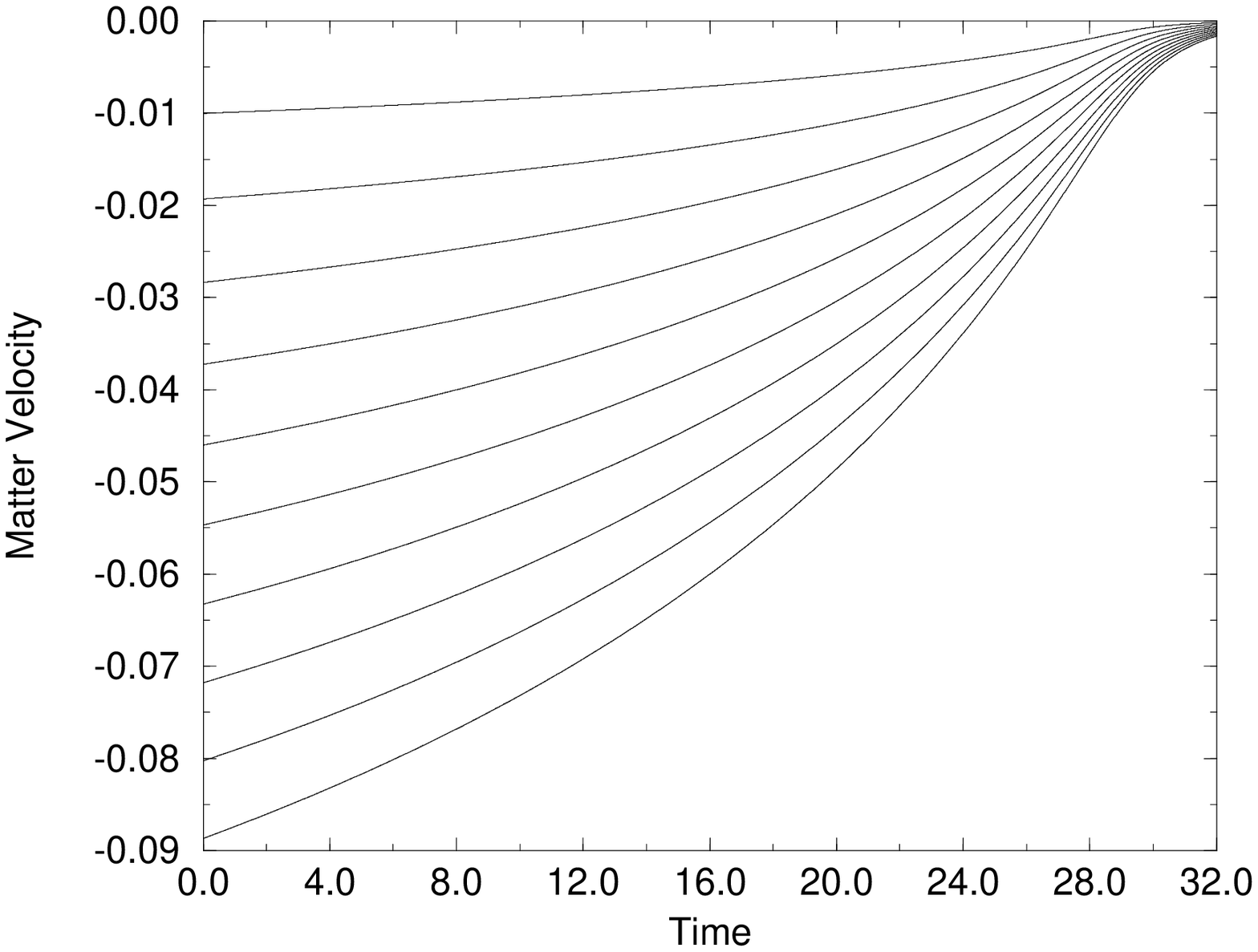}}
\caption{$dr/dt$ as a function of time, for different values of
 $r/a$: 0.1 (uppermost curve), 0.2, 0.3, 0.4, 0.5, 0.6, 0.7, 0.8, 0.9 and 1.0
 (lowermost curve).}
\label{fig:velocity}
\end{figure}

\begin{figure}
\centerline{\epsfxsize=6in\epsfbox{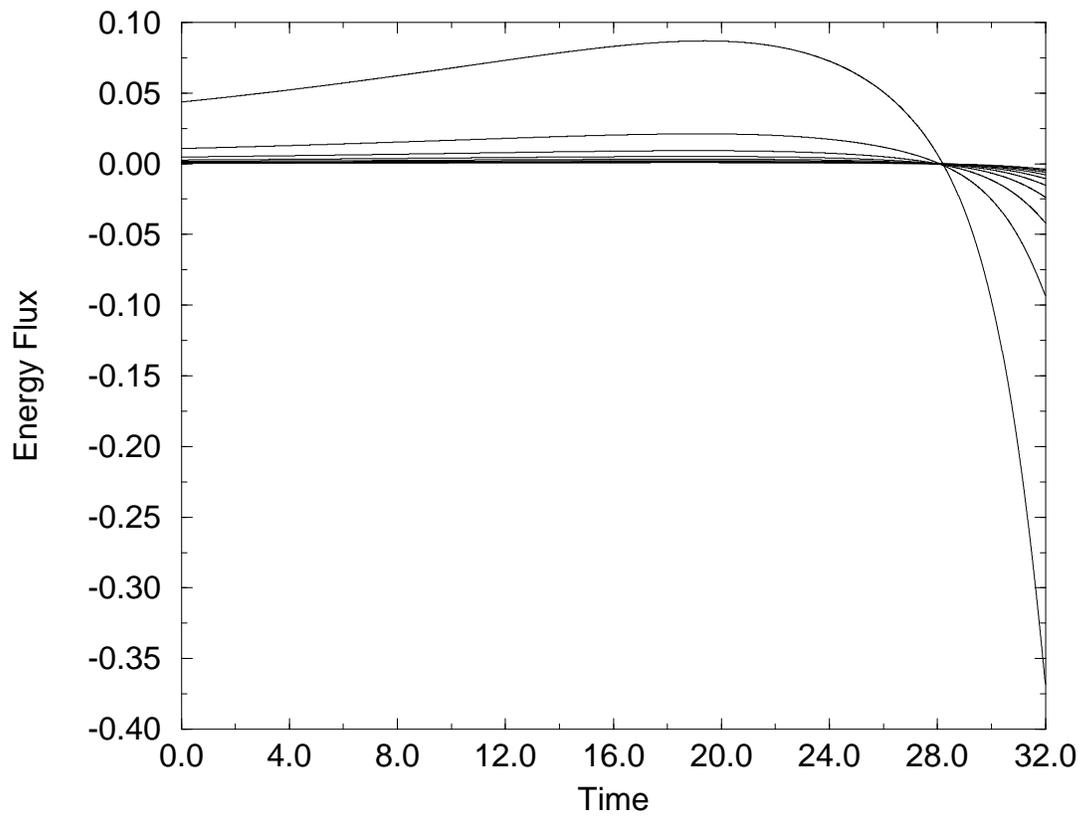}}
\caption{$\epsilon$ as a function of time, for different values of
 $r/a$: 0.1 (initially uppermost curve), 0.2, 0.3, 0.4, 0.5, 0.6, 0.7, 0.8,
 0.9 and 1.0 (initially lowermost curve).}
\label{fig:flux}
\end{figure}

\begin{figure}
\centerline{\epsfxsize=6in\epsfbox{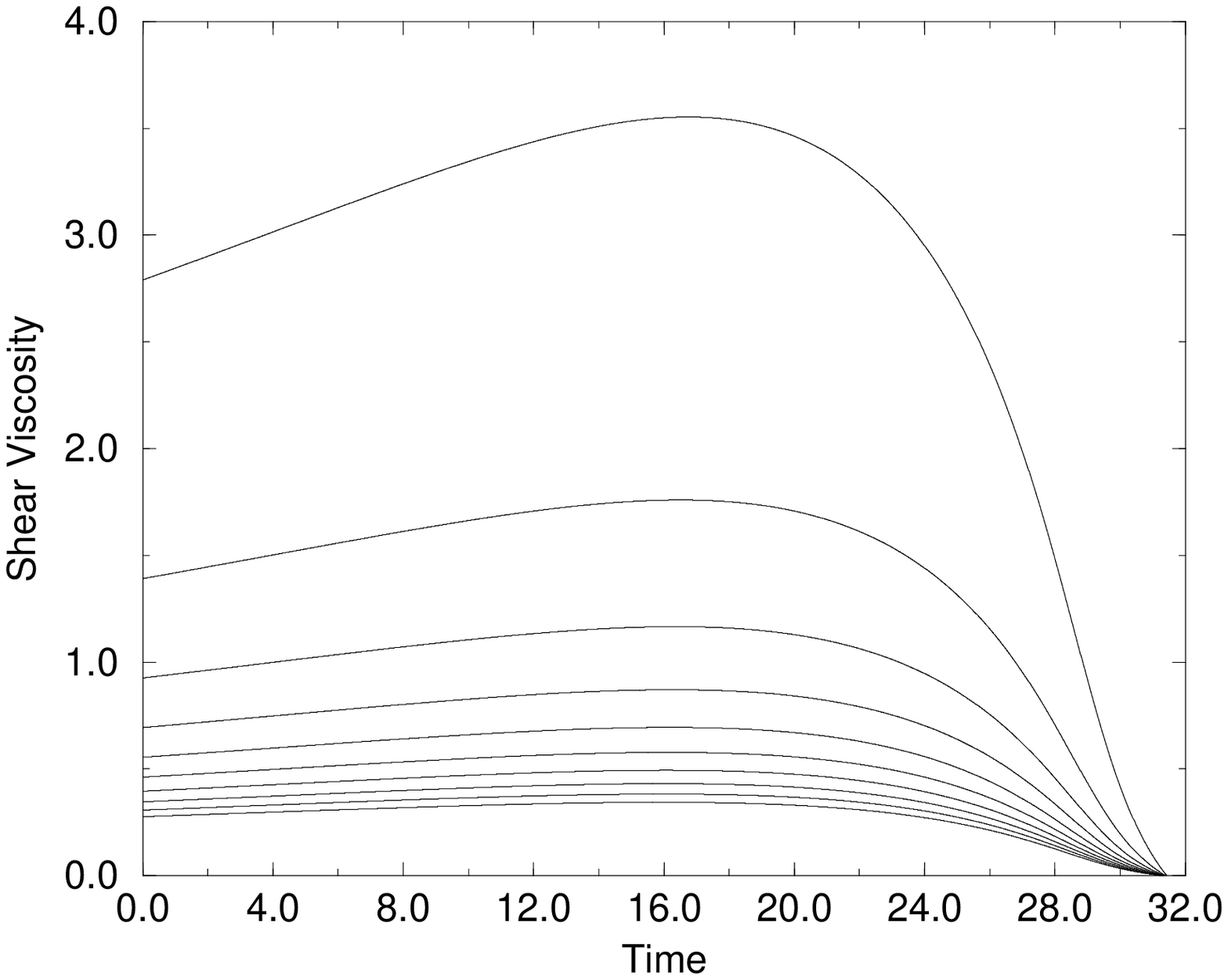}}
\caption{$\eta$ as a function of time, for different values of
 $r/a$: 0.1 (uppermost curve), 0.2, 0.3, 0.4, 0.5, 0.6, 0.7, 0.8, 0.9 and 
 1.0 (lowermost curve).}
\label{fig:viscosity}
\end{figure}

 We work out a ``tricky'' third scenario by
 ``forgetting'' the origin of parameter $l$, proposing that it depends
 on time in a very special way. If we imagine
 $N$ initial self--similar states which represents the history
 of the collapsing surface, the symmetry equations (\ref{eq:se3}) and
 (\ref{eq:se4}) are satisfied at every point of the space--time without
 taking into account the variation with time of $l$. Therefore, 
 we integrate numerically only
 equations (\ref{eq:first}) and (\ref{eq:second}), with
 (\ref{eq:luminosity}), (\ref{eq:omega}) and with $l=l(t)$. 
 Here we use standard Runge--Kutta
 (fourth order) methods and the initial conditions
$$
A(0)=3.255; F(0) \approx 0.3856
$$
Once the boundary evolution and its energetics are determined, we use (\ref{eq:mass})
 [or (\ref{eq:masa})] and (\ref{eq:nu}) to calculate the physical variables
 from the field equations. 
 Figures (\ref{fig:poverho})--(\ref{fig:viscosity})
 sketch the ratio $\tilde p / \rho$, $dr/dt$, $\epsilon$ and $\eta$, 
 respectively, for $k = (2\,) 10^{-3}$. These self--similar spheres
 do not have a barotropic equation of state [figure
 (\ref{fig:poverho})]. All shells evolve with decreasing collapsing velocities
 [figure \ref{fig:velocity})]. This behavior seems to be
 connected with the
 absorption of energy shown in figure (\ref{fig:flux}) in the late stage. Shear
 viscosity increases initially with collapse but later decreases with time
 on any shell.

\section{Conclusions}
~~~~We have assumed an additional symmetry to the space--time, homothetic
 motion, to generate non--static and simple solutions. These solutions
 were matched with the Vaidya one. We found that self--similar spheres
 with a barotropic equation of state ($\tilde p = 0$)
 are shear--free, this result is in complete accord with theoretical
 expectation \cite{op90}, \cite{wl88}\cite{hwe82}. 
 Other self--similar scenarios are possible as well if we 
 assume the evolution of the surface as a huge set of initial self--similar
 states. The shear viscosity profiles obtained in this work coincide qualitatively 
 surprisingly well with others calculated in a more realistic 
 framework \cite{m96}. 

\begin{center}
{\large Acknowledgments}
\end{center}
We benefited from research support by the Consejo de Investigaci\'on
under Grant CI-5-1001-0774/96 of the Universidad de Oriente and from
computer time made available from SUCI-UDO and CeCalCULA.

\newpage

\end{document}